\documentclass{article}
\usepackage{amssymb}

\newcommand{\bel}[1]{\begin{equation}\label{#1}}

\newcommand{\be}{\begin{equation}}
\newcommand{\ee}{\end{equation}}
\newcommand{\ba}{\begin{eqnarray}}
\newcommand{\ea}{\end{eqnarray}}
\newcommand{\rf}[1]{(\ref{#1})}
\newcommand{\bi}{\bibitem}

\newcommand{\tn}{$\vartheta$-neuron}

\newcommand{\ifn}{integrate-and-fire neuron}
 \newcommand{\st}{s^\star}

\begin{document}
\title{Temporal correlation based learning in neuron models} 

\vspace{2.5cm} 

\author{J\"urgen Jost\footnote{Max Planck Institute for Mathematics in the
  Sciences, Inselstr.22-26, 04103 Leipzig, Germany, jost@mis.mpg.de}
\footnote{Santa Fe 
  Institute, 1399 Hyde Park Road, Santa Fe, NM 87501, USA, jost@santafe.edu}} 
\maketitle 

\begin{abstract}We study a learning rule based upon the temporal correlation
  (weighted by a learning kernel) 
  between incoming spikes and the internal state of the postsynaptic
  neuron, building upon previous studies of spike timing dependent
  synaptic plasticity (\cite{KGvHW,KGvH1,vH}). Our learning rule for the synaptic weight $w_{ij}$ is
$$
\dot w_{ij}(t)= \epsilon
\int_{-\infty}^\infty \frac{1}{T_l} \int_{t-T_l}^t \sum_\mu
\delta(\tau+s-t_{j,\mu}) u(\tau) d\tau\ \Gamma(s)ds
$$
where the $t_{j,\mu}$ are the arrival times of spikes from the
presynaptic neuron $j$ and the function $u(t)$ describes the state of
the postsynaptic neuron $i$. Thus, the spike-triggered average contained in the inner integral is weighted by a kernel $\Gamma(s)$, the learning window,
positive for negative, negative for positive values of the time diffence  $s$ between post- and presynaptic activity. An antisymmetry assumption
  for the learning window 
  enables us to derive analytical expressions for a general class of
  neuron models and to study the changes in input-output relationships
  following from synaptic weight changes. This is a genuinely
  non-linear effect (\cite{SMA}).
\end{abstract}

\section{Introduction}
This paper deals with the question of  
incorporating correlation based learning mechanisms in formal neuron models. According to the so-called Hebb rule, such correlations are encoded by synaptic weights, and learning is considered as a mechanism implementing this. More precisely, the decisive feature of the mechanism of synaptic plasticity
is its response to temporal correlations in the input-output pattern
of the postsynaptic neuron. The pattern of STDP, spike timing
dependent synaptic plasticity, emerged both from profound theoretical investigations
of Gerstner, Kempter, van Hemmen, Wagner, and others
(\cite{KGvHW,KGvH1,KGvH2,LKvH,vH,XS}), inspired by 
discoveries about the processing of spatial information in the
auditory system of the barn owl, from detailed experimental studies
(see \cite{MLFS,Zh,BP,BP2}),
and from carefully set up computer simulations (\cite{SMA}). Put simply, the result is that a synapse
gets strengthened when a presynaptic input arrives shorthly before the
generation of a postsynaptic spike, and that it gets weakened if the
two events occur in reverse temporal order. This also solved one of
the problems for the implementation of Hebb's rule in older network models that
followed a continuous dynamics, namely that one needed an
additional mechanism that was either ad hoc or non-local to prevent
the synaptic strengths from growing without bounds (see \cite{AN}).\\
The theoretical analysis, however, ultimately needed
a neuron model that was linear in the sense that the neuron's
activity, spiking probability, or spike rate depended in a linear
manner on the input received. The resulting learning rule, while based
on the correlation between in- and output, then did not in turn affect
that correlation. The computer simulations of Song et al.\cite{SMA}, however,
demonstrated that such learning based on temporal correlations in a
non-linear neuron model could sharpen those correlations and thus make
the operation of the neuron more efficient in this sense. It is the
purpose of the present article to provide a framework within which
temporal correlation based learning and the resulting changes in those
correlation patterns can be analyzed for general
non-linear neuron models.\footnote{While we can treat non-linear neuron models,
however, the learning rule employed is itself linear in the sense that
it assumes a linear dependence of the
weight changes on some internal state function; that state function could be the firing frequency, but it could also be some non-linear function of it.} For this, we shall need to
make one crucial 
symmetry assumption about the shape of the learning window which, while
not in direct qualitative contrast with the neurobiological findings,
apparently is not strictly quantitatively valid. If we take as our state function the firing rate of the postsynaptic neuron (or some function of that firing rate), our model is a hybrid between a timing and a rate dependent model. Still, as argued in 
\cite{Sjo}, for capturing the full wealth of experimental findings, probably models are needed that include a more refined relationship between timings and rates. For those
reasons, our results 
can only be considered as a somewhat crude approximation of the
underlying neurobiological reality, but we hope that the elegance of
the theoretical principle will still provide us with useful
insights. It remains the task of deriving this principle from
information theoretic considerations.

\section{Temporal correlations and the learning rule}
We consider the synaptic strength $w_{ij}$ from a presynaptic neuron
$j$ to a postsynaptic one $i$. $i$ shall stay fixed over the entire
course of our analysis, and so the dependence on $i$ could well be
omitted from our symbols, but we find it useful to include it
nevertheless. By way of contrast, we shall need to compare the effects
of several presynaptic neurons, and so the index $j$ is
indispensable. The learning rule then is a differential equation for
$w_{ij}$ as a function of time $t$, depending on the temporal
correlation between pre- and postsynaptic activities, i.e., the ones of
the neurons $j$ and $i$.\\
The aim that the learning rule should depend on temporal correlations
between pre- and postsynaptic activity and the fact that the activity
and the learning dynamics take place on different time scales makes it
necessary to specify the relation between those time scales. Based
upon \cite{KGvHW,KGvH1}, it is carefully explained in \cite{vH} that
the time scale 
$T_l$ 
on which the synaptic weight changes are analyzed should separate the
time scale of the neural spikes and the typical interspike intervals
from the one where the effects of learning become visible. More
precisely, both an ``ergodic'' and an ``adiabatic'' hypothesis is
employed. The first one allows to average over randomness, that is,
over repeated trials, or, 
equivalently, over the time course of a single, sufficiently long,
trial whereas the second one allows to average over time, that is, to
work with quantities that can 
be assumed to be constant on a time interval of length $T_l$. Here, we
shall also employ the latter, the adiabatic hypothesis, but use the
former only in a somewhat weaker form.\\
The presynaptic activity is given by the spike train 
\bel{eq1}
\rho_j(t)=\sum_{\mu} \delta(t-t_{j,\mu})
\ee
produced by $j$. Here, the $t_{j,\mu}$ are the firing times of neuron
$j$ in the time period under consideration, and we use the standard
$\delta$-function formalism. For the postsynaptic neuron $i$, instead
of directly working with its spike train, we
employ a state function $u(t)$. Since this is fundamental for our
approach, we should carefully discuss the underlying reasoning:
\begin{itemize}
\item While a presynaptic spike is an event that cannot be further
  analyzed and has to be considered as externally caused, the activity
  and the firing pattern of the postsynaptic neuron depend on the
  inputs received from many other neurons, not only from the one
  forming the synapse under consideration. Therefore, it is appropriate to work
  with  some average for the total input of the neuron and to consider
  its state as the dynamical response averaged over the total input.
\item In contrast to the presynaptic spike, a postsynaptic spike
  should not be considered as an independent event. We rather need to
  employ a model for the dynamic activity of the neuron in response to
  its input, i.e., we need to suppose some deterministic relationship
  between input and output -- not necessarily between individual
  spikes, but for example between an incoming spike and a postsynaptic
  spiking probability. The effect of a single presynaptic spike may be
  very slight, but nevertheless it should make some definite
  contribution to the state of the postsynaptic neuron. That
  contribution may well depend on previous contributions from other
  presynaptic neurons, and on the present postsynaptic state
  itself, and this will then lead us to non-linear models. 
\end{itemize}
We shall discuss various possibilities for the state function $u$
below. At the moment, we need not specify it any further. The next
ingredient in our learning rule is  the shape $\Gamma(s)$ of the time
window that describes how the time difference $s$ between pre- and
postsynaptic activity influences the synaptic weight change. It is
generally agreed in the literature on STDP that 
both experimental evidence (see \cite{BP} for a review) and conceptual
reasoning (\cite{KGvHW,KGvH1,vH}) lead to the 
following qualitative behavior: 
\begin{enumerate}
\item $\Gamma(s)=0$ whenever the absolute value of the time difference
  $s$ between pre- and postsynaptic spike is large.
\item $\Gamma(s)>0$ whenever the presynaptic spike shortly precedes
  the postsynaptic one, i.e., when $s$ is negative and of small
  absolute value.\footnote{For an exception where there has been found an additional negative window for a certain range of negative $s$, see \cite{Nis}.} This effect has been called LTP (long term potentiation). This is interpreted as a manifestation of causality, 
  in the sense that the presynaptic spike can then be considered as
  contributing to the postsynaptic one. 
\item $\Gamma(s)<0$ whenever the presynaptic spike occurs shortly after
  the postsynaptic one, i.e., when $s$ is positive and of small
  absolute value. This effect is called LTD (long term depression).
\end{enumerate}
In addition, we shall need one further assumption on $\Gamma(s)$:
\begin{itemize}
\item[4. ] $\Gamma(s)$ is balanced in the sense that
\bel{eq2}
\int_{-\infty}^\infty \Gamma(s) ds=0.
\ee
This assumption will be needed subsequently to convert an integral of a neuronal state function $u$ against the time window function $\Gamma$ into an integral of a differences of states against $\Gamma$ restricted to positive arguments, in other words for evaluating and weighting the effect that an input causes on the state of the postsynaptic neuron. In our analysis below, we shall mostly work with a state function that represents a firing probablity of the postsynaptic neuron. In that situation, the function $\Gamma$ can be interpreted as the neurophysiological learning window, and it then becomes an issue whether that balancing condition is experimentally supported. We shall summarize some of the findings in this direction shortly. Our framework below, however, also allows us to consider state functions that do not directly correspond to firing probabilities, but perhaps some other internal quantities, or that subject such a firing probability to some non-linear transformation. Under such more liberal conditions, the balancing assumption employed here is still compatible with a non-zero integral of the neurophysiological learning window.\\ 
While, returning to spike probabilities, this balancing assumption does not seem too far from the neurophysiological findings,
it is not strictly supported by them. At least, at present the
available data seem to be mixed on this issue (see the discussion and
the references in \cite{SMA}). 
$\int_{-\infty}^\infty \Gamma(s) ds$ has not always been found to be
strictly 0, but at least it seems to be rather small in most
situations. Recent experimental results that report a non-zero integral are, for example, described in \cite{Deb,Feld,FrDan,Sjo}\footnote{It is not clear to the present author, however, how accurate the estimates for the integral are. For example, in \cite{BP}, a negative integral is computed on the basis of fitting the data by exponential functions, and the negative integral then results from a longer decay time of LTD. That the data are described well by an exponential function, however, is not entirely obvious to the present author, and fitting them with a different type of function might well lead to a different value and sign for the integral.}, typically with a longer window and a slight dominance for LTD, whereas those finding similar time frames for strenghtening and weakening  report approximate equality between the two effects or a slight dominance for LTP \cite{MLFS,BP3}. In any case, since the detailed electrochemical processes
at 
a synapse 
before and after reception of a spike are different, there cannot be
any direct biophysical reason for \rf{eq2}. At most, there may be a
very indirect reason for an  approximate  validity of
\rf{eq2}. Namely, if that relation should turn out to be most
advantageous for the processing of information in neuronal systems, it
might be implemented by the forces of evolution. This, however, may
seem a little far-fetched in the present context, and so, this
assumption can be defended here only on the basis of its analytical
utility. The condition  \rf{eq2}, $\int_{-\infty}^{\infty} \Gamma(s)=0$, can,
however, be also considered as a normalization that prevents the
synaptic weights from growing without 
  bounds when subjected to Hebb type learning rules. As argued in
  \cite{SMA,AN}, however, for that purpose a (small) negative value of
  that integral would be better, in order to weaken inputs that do not
  consistently contribute to postsynaptic firing, but only by chance
  occur at about the time of that firing. Uncorrelated pre- and
  postsynaptic activities would then induce a weakening of the
  synapse. By way of contrast, under our assumption, if we take as our state variable for the postsynaptic neuron its firing rate, statistically
  independent pre- and postsynaptic spike trains produce no weight
  change. However, since we may use a non-linear function of the firing rate, we still possess a certain flexibility here. The really serious issue is that our symmetry assumption prevents us from making a rate-dependent distinction between LTP and LTD. Such a distinction, i.e., that at low rates, LTD dominates, whereas at higher rates, LTP does, has been discussed in \cite{Sjo}. Also, these authors have found that a model that includes rate, timing, and cooperativity captures some experimental findings better than one that solely depends on the relative timing of pre- and postsynaptic spikes. In their model, only nearest spike interactions count, and LTP wins over LTD. In partial contrast to this, Froemke and Dan\cite{FrDan} found that synaptic modification depends not only on the relative timing of pre- and postsynaptic spikes, but also on the spiking patterns within each neuron. In particular, the first spike within a burst is found to be the dominant one. \\
As
  a consequence of our symmetry assumption
   \rf{eq2}, there exists a transformation 
$s^\star(s)$ of the positive reals onto themselves such that for $s>0$,
$\Gamma(-s^\star(s))\frac{ds^\star}{ds} = - \Gamma(s)$, and
consequently, for functions $v$ (assuming natural integrability
assumptions),
\bel{eq2a}
\int_{-\infty}^0 v(\sigma) \Gamma(\sigma)\ d\sigma = \int_0^\infty
v(-s^\star(s))(-\Gamma(s))\ ds.
\ee
(The formally simplest case is the one where $\Gamma$ is an odd
symmetric function, in the sense that $\Gamma(-s)=-\Gamma(s)$ for all
$s$; in that case, of course $\st(s)=s$.) The point here is that in general the time courses realizing LTD  and LTP  are different (see \cite{KD} for a recent survey of the molecular mechanisms underlying these processes).  By the symmetry assumption on $\Gamma$ the total, time-integrated effects of the two processes are of the same magnitude, but stretched differently over time. The function $s^\star(s)$ rescales the latter process to make it symmetric to the former.
\end{itemize}
The idea for our learning rule is now quite simple: We  convolve an input-output correlation spike-triggered average with the learning window function $\Gamma$ to obtain 
 a differential equation for the
synaptic weight $w_{ij}$:
\bel{eq3}
\dot w_{ij}= \epsilon
\int_{-\infty}^\infty \frac{1}{T_l} \int_{t-T_l}^t \sum_\mu
\delta(\tau+s-t_{j,\mu}) u(\tau) d\tau\ \Gamma(s)ds\\
\ee
where the sum extends over all spikes between $t$ and $t-T_l$ (and
where we neglect the discontinuities as a function of $t$ when a spike
happens to fall precisely on one of these boundary points, because
$T_l$ is supposed to be much larger than the typical interspike
interval). The spike-triggered average occurring here, that is, the inner integral, is known in the literature as stimulus-response correlation or peri-stimulus time histogram. In other words, we stipulate that the synaptic weight change is proportional to an input-output correlation weighted by the kernel $\Gamma$. The proportionality factor $\epsilon$ is supposed to be  small, to make
learning into a slow process that shows significant effects only on a
time scale larger than $T_l$, as in \cite{KGvHW,KGvH1,vH}. \rf{eq3} is 
\bel{3a}
=\epsilon \frac{1}{T_l} \sum_\mu \int_{t_{j,\mu}-t}^{t_{j,\mu}-t+T_l}
u(t_{j,\mu}-s)\ \Gamma(s)ds.
\end{equation} 
Following \cite{vH}, we may use the adiabatic hypothesis to extend the integration boundaries for $s$ to $\pm \infty$. Namely, when $t_{j,\mu}=t-T_l + xT_l$ for $0<x<1$, $s$ in the integral varies between $(x-1)T_l$ and $xT_l$, and for any given $x$ and sufficiently large $T_l$, the integration bounds can be taken as $\pm \infty$ because $\Gamma(s)=0$ for $|s|>>1$. Thus, we may write
\begin{eqnarray}
\label{eq4}
\dot w_{ij}&=&\epsilon \int_{-\infty}^\infty \frac{1}{T_l} \sum_\mu
u(t_{j,\mu}-s)\ \Gamma(s)ds\\ 
\nonumber
&=&\epsilon \frac{1}{T_l} \sum_\mu \int_0^\infty
(u(t_{j,\mu}+s^\star(s))-u(t_{j,\mu}-s))(-\Gamma(s))ds 
\end{eqnarray}
because of the symmetry assumption 4 above, and this
 receives a positive contribution if $u$ is larger after the incoming spike $t_{j,\mu}$
than before
\bel{eq5}
=\epsilon \frac{1}{T_l} \sum_\mu \int_0^\infty \int_{-s}^{s^\star(s)} \dot
u(t_{j,\mu}+\tau) d\tau (-\Gamma(s))ds 
\ee
which receives a positive contribution if $u$ is increasing between $t_{j,\mu}-s$ and
$t_{j,\mu}+\st$.\\

\section{Neuron models}
In order to gain some preliminary understanding of the learning rule \rf{eq3}, we briefly consider the case where the state function $u$ is also given by a spike sum, $u(\tau)=\sum_\nu \delta(\tau-t_{i,\nu})$. Inserting this into \rf{eq3} yields
\begin{eqnarray}
\label{eq20}
\nonumber
\dot w_{ij}&=& \epsilon
\int_{-\infty}^\infty \frac{1}{T_l} \int_{t-T_l}^t \sum_{\mu,\nu}
\delta(\tau+s-t_{j,\mu}) \delta(\tau-t_{i,\nu} ) d\tau\ \Gamma(s)ds\\
\nonumber
&=& \epsilon
\int_{-\infty}^\infty \frac{1}{T_l}  \sum_{\mu,\nu}
\delta(t_{j,\mu}-t_{i,\nu}-s)\ \Gamma(s)ds\\
&=& \epsilon \frac{1}{T_l}\sum_{\mu,\nu} \Gamma(t_{j,\mu}-t_{i,\nu}).
\end{eqnarray}
Thus, the learning rule changes the synaptic weight by a weighted sum of spike time differences, as it should. \\
In order to gain deeper insights, however, we need to assume a systematic relationship between input and output, and the approach taken here is that the state of the output neuron $i$ is described by a state function $u$ whose value then  is determined by the inputs received. In the preceding example, we could insert a functional relationship that determines the difference in spike times $t_{j,\mu}-t_{i,\nu}$ as a function of the synaptic input received, and thus, assuming the presynaptic activity as given, of the synaptic weight $w_{ij}$. This will then lead to a differential equation for the weight change in terms of that -- and perhaps other -- synaptic weights. Depending on that functional relationship, the weights in the model may then become unbounded or, which would be more pleasing, converge to some stationary value. The mechanism for the latter is that an increased weight might decrease the spike difference to 0 so that then there will be no further increase since $\Gamma(0)=0$. The analysis is rather straightforward and therefore omitted. We point out, however, that a vanishing, or even negative, integral of the learning kernel $\Gamma$ does not automatically imply that the synaptic weights stay bounded in such models. Namely, once a certain weight $w_{ij}$ has grown so large that any presynaptic input triggers a postsynaptic spike, then, unless that spike occurs without any time delay simultaneously with the input, the weight will keep growing, unless some input from a different synapse causes spikes shortly before the present input. Conversely, however, once our weight is so large, it will then  exert that negative effect on other synapses rather than being dampened by those.\\
We now turn to a more systematic analysis of how assumptions about the functional relationship between the input and the state $u$ of our postsynaptic neuron drive the weight changes in our setting. In fact,  we can insert any neuron
model in \rf{eq5} that is given either by an explicit expression for
the state function  $u$ or in the form of a differential equation
for that state function. We start with the first possibility and
consider a model where the state of our neuron $i$ is computed as
\bel{eq21}
u=f(\sum_k w_{ik} \sum_\nu \kappa(t-t_{k,\nu})).
\ee
Here, the $t_{k,\nu}$ are the spike times of the incoming neurons $k$,
and $\kappa$ is a synaptic transfer function. For causality, $\kappa$
needs to satisfy the requirement that $\kappa(\tau)=0$ for $\tau \le
0$, i.e., an incoming spike can only influence the state of neuron $i$
after it occurred. Reasonable choices for $\kappa$ that have been
discussed in the literature are (see \cite{GK} for a general treatment):\\
1) A function that sets in sharply upon the arrival of a spike at time
$t_{j,\mu}$ and
then decays exponentially:
\bel{eq22}
\kappa_0(t-t_{j,\mu}):=
\exp(-\frac{t-t_{j,\mu}}{\tau})H(t-t_{j,\mu}),
\ee
with the usual Heavyside function $H$.
The derivative of this function is given by
\bel{eq23}
\dot \kappa_0(t-t_{j,\mu})=
\delta(t-t_{j,\mu})-\frac{1}{\tau}\exp(-\frac{t-t_{j,\mu}}{\tau})H(t-t_{j,\mu})
.
\ee
2) The so-called $\alpha$-function that starts linearly upon arrival
of a spike and again decays exponentially:
\bel{eq24}
\kappa_1(t-t_{j,\mu}):=
\frac{t-t_{j,\mu}}{\tau}\exp(-\frac{t-t_{j,\mu}}{\tau})H(t-t_{j,\mu}).
\ee
3)
Both $\kappa_0$ and $\kappa_1$ have the disadavantage that they do not
return to 0 in finite time. This means that an incoming spike will
have some effect forever, even though this effect of course decays
exponentially. For that reason, one might prefer a decay to 0 in
finite time, for example a linear one:
\bel{eq25}
\kappa_2:= 1-\frac{t-t_{j,\mu}}{\tau} \mbox{ for } 0 \le
\frac{t-t_{j,\mu}}{\tau} \le 1
\ee
and 0 otherwise.\\
The function $f$ translates the synaptically weighted input sum into
some activation of neuron $i$; it could stand for its firing rate,
probability, or propensity. In those cases, one would assume that it
is monotonically increasing, i.e., has a positive derivative.\\ 
With this neuron model, 
\rf{eq4} then becomes
\ba
\label{eq26}
\epsilon \int_0^{\infty}\frac{1}{T_l} &\sum_\mu  &(f(\sum_k w_{ik} \sum_\nu
\kappa(t_{j,\mu}-t_{k,\nu}-s))\\
\nonumber
&-& f(\sum_k w_{ik} \sum_\nu
\kappa(t_{j,\mu}-t_{k,\nu}+s^\star(s))))\ \Gamma(s) ds.
\ea
Since $\Gamma(s)<0$ for $s>0$, the contribution to this expression at
$s>0$ is positive if $f(\sum_k w_{ik} \sum_\nu 
\kappa(t_{j,\mu}-t_{k,\nu}+\st))>f(\sum_k w_{ik} \sum_\nu
\kappa(t_{j,\mu}-t_{k,\nu}-s))$, i.e., if the state of $i$ is larger
after the spike at $t_{j,\mu}$ than before. We recall that
$\kappa(\sigma)$ vanishes for $\sigma<0$, and that (for the choices $\kappa_0$ or $\kappa_2$, say) it is a decreasing
function of $\sigma \ge 0$. Thus, if $t_{j,\mu} - t_{k,\nu} +\st \ge 0 >
t_{j,\mu} - t_{k,\nu}-s$, i.e., if $t_{j,\mu} -s < t_{k,\nu} \le
t_{j,\mu} +\st$, we can expect a positive contribution, whereas if the
spike of $k$ comes too early, $t_{k,\nu} \le
t_{j,\mu} -s$, we obtain a negative contribution. In other words, for
a spike of neuron $j$ to strengthen the synapse to neuron $i$, it
needs to be good at anticipating or ``predicting'' other incoming
spikes that in turn increase the state of $i$ and thus its likelihood
to fire. We thus confirm the conclusion of the previous investigations
\cite{KGvHW,KGvH1,vH,GK}.\\
We now use one of the transfer functions $\kappa_0$ or $\kappa_2$ for
$\kappa$; the modifications required for the choice $\kappa_1$ will be
obvious. We obtain from \rf{eq5}
\begin{eqnarray}
\label{eq27}
\nonumber
 &\epsilon& \int_0^{\infty}\frac{1}{T_l} \int_{-s}^{s^\star} \sum_\mu
\frac{df}{d\tau}(\sum_k w_{ik} \sum_\nu 
\kappa(t_{j,\mu}-t_{k,\nu}-\tau))\\
=&\epsilon& \int_0^{\infty}\frac{1}{T_l} \int_{-s}^{s^\star} \sum_\mu
f^\prime(\sum_k 
w_{ik} \sum_\nu 
\kappa(t_{j,\mu}-t_{k,\nu}-\tau))\\
\nonumber
& &\sum_l w_{il} \sum_\rho
(-\delta(t_{j,\mu}-t_{l,\rho}-\tau) - \dot
\kappa(t_{j,\mu}-t_{l,\rho}-\tau))d\tau\ \Gamma(s) ds 
\end{eqnarray}
(where in contrast to \rf{eq23}, the derivative $\dot\kappa$ stands
only for the smooth part) 
which in turn equals
\begin{eqnarray}
\label{eq28}
& &\epsilon \sum_\mu \frac{1}{T_l} \int_0^{\infty}  \sum_l w_{il}
\sum_{\rho}(f^\prime(\sum_k w_{ik} \sum_\nu 
\kappa(t_{l,\rho}-t_{k,\nu}))\\
\nonumber
 &+& \int_{-s}^{s^\star}   f^\prime(\sum_k
w_{ik} \sum_\nu 
\kappa(t_{j,\mu}-t_{k,\nu}-\tau)) \dot
\kappa(t_{j,\mu}-t_{l,\rho}-\tau)d\tau)\ (-\Gamma(s)) ds.
\end{eqnarray}
Note that we have shifted the minus sign to the $\Gamma$-term. The
first term here counts the number of spikes of $j$ with 
appropriate weights. In the second one, since
$\kappa$ is decreasing, $\dot \kappa$ yields a negative factor.\\\\
We now turn to neuron models where the derivative of the state
function $u$ is given in terms of the input and some inherent dynamic
features. An example is the leaky \ifn\ that is described by
\bel{eq29}
 \frac{du(t)}{dt}= - c_1 u(t) + \sum_k
w_{ik} \sum_\nu 
\kappa(t-t_{k,\nu}) -  u_{th} \sum_{t_{i,\lambda}}
  \delta(t-t_{i,\lambda}) 
\ee
for some constant $c_1$ that determines the time scale of the
decay.\footnote{Perhaps this example is not so good for the present
  setting, however, because in the \ifn\ model, $u$ denotes a membrane
  voltage and cannot be interpreted as a firing probability.} $u_{th}$
is the value of the spiking threshold, and the 
$t_{i,\lambda}$ are the spiking times of our neuron $i$. We may then
insert \rf{eq29} into \rf{eq5} to obtain the learning rule for the
leaky \ifn . The analysis can then proceed as for the above state
function $f$ where we encountered the derivative of the transfer
function $\kappa_0$ or $\kappa_2$, except that the signs get reversed
because 
an output spike gives a
negative instead of a positive contribution. The model is not entirely
satisfactory because the output spike time is not explicitly
determined as a function of the input. Also, the discontinuous
resetting of $u$ to 0 upon the emission of a spike cannot directly be
approximated by a smooth dynamics for the scalar variable $u$ when $u$
is given as a solution of an ODE, since a solution of such an ODE can
only exhibit monotonic behavior when continuous. We can overcome this
problem, however, by introducing an internal state variable
$\vartheta$ that takes its values in the unit circle instead of the
interval $[0,u_{th}]$ so that $u=u(\vartheta)$ is a 2-1 function. An
example of such a model is the \tn\ introduced and studied by
Ermentrout and Gutkin \cite{Er2,GuEr} (see also \cite{GK} for a
general discussion). We can set things up in such a manner
that $\vartheta=0$ corresponds to the rest point $u=0$ and $\vartheta=\pi$
to the firing point $u=u_{th}$ of our neuron. The neuron model then consists in
expressing the time derivative $\dot u$ of $u=u(\vartheta(t))$ as a
function of $\vartheta(t)$ and the input $I$ ($=\sum_k
w_{ik} \sum_\nu 
\kappa(t-t_{k,\nu})$, as always),
\bel{eq30}
\dot u = \frac{du}{d\vartheta} \dot\vartheta = \Phi(\vartheta(t), I(t)).
\ee
We can then describe the generation and emission of a spike in a
continuous manner. What is essentially needed is that the 
derivative $\frac{du}{d\vartheta}$ is positive in
the upswing phase $0<\vartheta<\pi$ and negative in the downswing
phase $\pi<\vartheta<2\pi$ (where, of course, $2\pi$ is periodically
identified with 0). The precise shape of $\Phi$ will, of course,
depend on the biophysical model employed, but this is not our present
concern. Qualitatively, the generation and emission of a spike and the
subsequent resetting of a neuron then is described by a hump function
$v(t-t_{i,\lambda})$ that can be made narrow, for example $v(\tau)>0$
precisely for $|\tau|\le \delta_0$ for some small $\delta_0 >0$, with
$v(0)=u_{th}$. We then have the conservation
$\int_{-\delta_0}^{\delta_0} \dot v(\tau) d\tau =0$. Thus, a spike that
occurs before the input at time $t_{j,\mu}$ then leads to a negative
contribution in \rf{eq5} for all $s$ with
$t_{i,\lambda}-t_{j,\mu}-\delta_0<-s<t_{i,\lambda}-t_{j,\mu}+\delta_0$
and to a vanishing one for other values of $s$.\footnote{Let us point out that $s$ is an integration variable, and so, this does not require that input and output spike appear virtually at the same time to have an effect. Rather, the function $\Gamma$ is only evaluated at a narrow range of its arguments $s$, the extreme case of course being the one described in \rf{eq20}.} Conversely, if
$t_{j,\mu}<t_{i,\lambda}$, we get a positive contribution for
$t_{i,\lambda}-t_{j,\mu}-\delta_0<s<t_{i,\lambda}-t_{j,\mu}+\delta_0$
and a vanishing one for other $s$.\\
For the subsequent analysis, a model of the qualitative form
\bel{eq31}
\dot u = F(\vartheta) + G(\vartheta)\sum_k w_{ik} \sum_\nu \kappa(t-t_{k,\nu})
\ee
is particularly useful. Here, $F$ can stand for a leakage term that
then is always non-positive, whereas $G(\vartheta)$ should be positive
before the spike (i.e., for $0<\vartheta<\pi$) and negative
afterwards so that $u$ can return to its rest value. A spike is then
built up when the r.h.s. of \rf{eq31} is positive for a sufficiently
long time. For that model,
\rf{eq5} becomes
\begin{eqnarray}
\label{eq32}
&\dot w_{ij}(t)& =\\
\nonumber
&\epsilon& \sum_\mu \int_0^{\infty}\frac{1}{T_l} 
\int_{-s}^{s^\star} (F( \vartheta(t_{j,\mu}+\tau))\\
\nonumber 
& &+G(\vartheta(t_{j,\mu}+\tau)) \sum_k w_{ik} \sum_\nu
\kappa(t_{j,\mu}-t_{k,\nu}+\tau) d\tau)\ (-\Gamma(s)) ds. 
\end{eqnarray}
The first sum here again extends over all spikes that come in from neuron $j$ with $t-T_l \le t_{j,\mu} \le t$. It may then happen that the argument $t_{j,\mu} +\tau$ in the integral is larger than $t$. This violates causality. However, as argued in Section 4 of \cite{vH}, those arguments are negligible because the learning window can be assumed much smaller than $T_l$. \\ 
Let us assume for the moment for the sake of the discussion that the
state variable $\vartheta$, being the result of many incoming spikes,
varies so slowly that we may replace $\vartheta(t+\tau)$ by
$\vartheta(t)$ for $-s \le \tau \le \st$. Then,  the contribution of the
spike at $t_{j,\nu}$ is positive when it occurs for $0 < \vartheta <
\pi$, since the integrand is positive in that case, i.e., when the neuron
$i$ increases its state variable towards 
the firing point, and negative in the interval corresponding to
returning to rest after firing. The first term again
counts the incoming spikes of $j$, weighted with the value of
$F(\vartheta)$. Turning to the 
second term, and using the transfer function $\kappa_0$ or $\kappa_2$,
by the properties of
$\kappa$ again, 
that contribution for the spike of neuron $k$ at $t_{k,\nu}$ is
strongest in absolute value for $t_{j,\mu}= t_{k,\nu} +s$ because it
can then exert its effect during the whole interval $[-s,\st]$. Thus, as a
consequence of the decay of $\kappa$ after the initial pulse, it is
best for the spike of neuron $j$ to 
occur shortly after the one of neuron $k$, provided this happens at a
time when $i$ is responsive to incoming spikes. If $\kappa=\kappa_1$
 instead, then the optimal delay of the spike of $j$
after the one of $k$ is even longer as the effect of the spike of $k$
exhibits itself most strongly only after some delay. Returning to
$\kappa_0$ or $\kappa_2$, the effect of the spike $t_{k,\nu}$ is small, if it
occurred too early since then its contribution during the interval
$[t_{j,\mu}-s,t_{j,\mu}+\st]$ is weak, while it is also small if it
comes too late because then its influence starts too late. If $\kappa$
decays to 0 very rapidly, then it does not make that much of a
difference anymore at which point in the interval
$[t_{j,\mu}-s,t_{j,\mu}+\st]$ the spike of $k$ occurs. In any case,
because it is still somewhat better for a spike of $k$ to occur
slightly before than after the one of $j$, we cannot conclude in the
present model that it is advantageous for $j$ to anticipate or predict
the spikes of other incoming neurons, but rather to exploit the
effects of those spikes and to bring $i$ even closer to firing. In any
case, a spike of $j$ leads to an increase of the weight $w_{ij}$ when
it occurs at a time when the state of $i$ increases from the resting
to the firing value, and to a decrease otherwise.\\
These conclusions do not change significantly if we no longer replace $\vartheta(t+\tau)$ by
$\vartheta(t)$ for $-s \le \tau \le \st$. The term $F( \vartheta(t_{j,\mu}+\tau))$ is again straightforward to analyze. In the second term, the quantity $\sum_k w_{ik} \sum_\nu
\kappa(t_{j,\mu}-t_{k,\nu}+\tau)$ is now multiplied by the varying term $G(\vartheta(t_{j,\mu}+\tau))$ which we assume here positive when $t_{j,\mu}+\tau$ is smaller than the firing time $t_i$, and negative afterwards. When then the incoming spike at $t_{j,\mu}$ occurs before the firing time, the term $\kappa(t_{j,\mu}-t_{k,\nu}+\tau)$ then receives a positive factor as long as $t_{k,\nu}<t_{j,\mu}+\tau <t_i$, and a negative one afterwards. If for example $t_{j,\mu}=t_i$, then spikes of $k$ at  $t_{k,\nu}<t_{j,\mu}$ lead to a positive effect as long as $\tau <0$, and to a negative, albeit smaller one, afterwards. Spikes at $t_{k,\nu}>t_{j,\mu}$ have a negative effect for all $\tau$, although over a shorter span of time. While the precise contributions will depend on the detailed shape of the learning window $\Gamma$, as well as on the actual spike train $t_{k,\nu}$, the positive and negative effects should about balance each other for $t_{j,\mu}=t_i$, while if for example $t_{j,\mu}>t_i$, this balance will be shifted towards the negative ones. \\

\section{Input-output relationships}
In the preceding, we have investigated how the temporal relationships
(correlations weighted with a temporal kernel)  
between pre- and postsynaptic activity drive the learning process as
incorporated in synaptic weight changes. We now address the question
of how in turn these correlation patterns change as the result of
learning. In order to study this, we should separate the effects of the
learning dynamics from the ones of the activity dynamics. As learning
leads to changes in the dynamic behavior, we should assume that we
start from a situation where without learning, i.e., when considering
solely the activity dynamics and assume that the synaptic weights are
constant, the neuron responds to stationary input in a stationary
manner. In other words, without learning, the activity dynamics should
always reach some state where the neuron responds to the same input in
the same manner.  In particular, without learning, the
relationship between input and output should be constant in time. This
assumption then allows us to study the effects of learning on this
relationship. We proceed to do so. The relationship between input and
output determining the synaptic weight change in our learning rule was computed or estimated in \rf{eq4}, \rf{eq5} by 
\begin{eqnarray}
\label{eq41}
\nonumber
C_{{io}}(t):&=&\int_{-\infty}^\infty \frac{1}{T_l} \int_{t-T_l}^t \sum_\mu
\delta(\tau+s-t_{j,\mu}) u(\tau) d\tau\ \Gamma(s)ds\\
\nonumber
&=& \int_{-\infty}^\infty \frac{1}{T_l} \sum_\mu
u(t_{j,\mu}-s)\ \Gamma(s)ds\\
&=& \sum_\mu \int_0^\infty \frac{1}{T_l} \int_{-s}^{s^\star(s)} \dot
u(t_{j,\mu}+\tau) d\tau (-\Gamma(s))ds,
\end{eqnarray}
and we now wish to compute how this expression changes in time in
response to learning. We obtain
\begin{eqnarray}
\label{eq42}
\nonumber
& &C_{{io}}(t)-C_{{io}}(t-h)\\
\nonumber
&=&\sum_\mu \int_0^\infty \frac{1}{T_l}\int_{-s}^{\st} (\dot
u(t_{j,\mu}+\tau) - \dot u(t_{j,\mu}-h+\tau))d\tau (-\Gamma(s))ds\\
&=&\sum_\mu \int_0^\infty \frac{1}{T_l}\int_{-s}^{\st} \int_{-h}^0 \ddot
u(t_{j,\mu}+\tau+\sigma) d\sigma d\tau\ (-\Gamma(s))ds.
\end{eqnarray}
Letting $h$ tend to 0, we obtain
\bel{eq42aa}
\dot C_{io}(t)=\sum_\mu \int_0^\infty \frac{1}{T_l}\int_{-s}^{\st}  \ddot
u(t_{j,\mu}+\tau)  d\tau\ (-\Gamma(s))ds.
\end{equation}
Without learning, by our stationarity hypothesis, this expression
should vanish. Looking for example at the second to last line in \rf{eq42}, without
learning, the integral over the derivative of $u$ should be invariant
under a time shift of the argument, here by the amount
$h$. 
This means that for the above integral over $\ddot u$, we only need to
investigate how the integral over the derivative of $u$ changes by the
learning dynamics when increasing the time by $h$. We study this for
the model \rf{eq31},
\bel{eq42a}
\dot u = F(\vartheta) + G(\vartheta)\sum_k w_{ik} \sum_\nu
\kappa(t-t_{k,\nu}). 
\ee
In order to simplify our notation, we abbreviate $\sum_k w_{ik} \sum_\nu
\kappa(t-t_{k,\nu})$ as $w\kappa$; thus $w$ stands for the synaptic
weight under consideration. When subjected to a variation $\dot w$ of
$w$, \rf{eq42a} varies by
\bel{eq42b}
\frac{d}{dw} \dot u \ \dot w = (F^\prime(\vartheta)
\frac{d\vartheta}{dw} + G^\prime(\vartheta)\frac{d\vartheta}{dw}
w\kappa + G(\vartheta)\kappa) \dot w
\ee
(here, $^\prime$ denotes a derivative with respect to
$\vartheta$.)\footnote{Since we shall study effects on the time scale
  $T_l$ which is independent of $\epsilon$, the linearization
  performed here constitutes a valid approximation for small enough
  $\epsilon$.} 
For the sake of the discussion, we separate the indirect contribution
coming from the variation of the state variable $\vartheta$ in
response to the change of the weight parameter $w$ from the direct
effect of this weight change on $\dot u$ as given by the last term in
\rf{eq42b}. The latter is 
\bel{eq43}
G(\vartheta)\sum_k \dot w_{ik}(t) \sum_\nu \kappa(t-t_{k,\nu}).
\ee
Inserting this into \rf{eq42}, we then obtain for the effect of this
term on the weighted 
input-output 
correlation 
\ba
\label{eq44}
& &\sum_\mu \int_0^{\infty}\frac{1}{T_l} 
\int_{-s}^{\st} 
G(\vartheta(t_{j,\mu}+\tau)) \\
\nonumber
& &\sum_k \dot
w_{ik}(t_{j,\mu}+\tau) \sum_\nu 
\kappa(t_{j,\mu}-t_{k,\nu}+\tau)  d\tau\ (-\Gamma(s)) ds .
\end{eqnarray}
From \rf{eq32}, we obtain a 
coarse estimate for this 
temporal change then as the average sign of
\bel{eq45}
(F(\vartheta(t))+G(\vartheta(t))I(t))\ G(\vartheta(t))
\ee
where $I(t)$ is an abbreviation for the input. Since the first factor
must be positive when a spike is being built up and the second factor
must then also be positive so that the input contributes towards the
spike, their product then is also positive during that phase. During
the resetting phase, $F$ and $G$ are typically both negative, and so,
we obtain again a positive contribution to the change of the weighted 
correlation. Thus, in this expression, a non-linear dependence of the change of
state of the neuron on the present value of that state itself leads to
a sharpening of the weighted input-output correlations through our spike timing
dependent learning rule.\\
These findings remain valid when we look at the precise formula in place of \rf{eq45}. That formula results from inserting \rf{eq32} into \rf{eq44}:
\begin{eqnarray}
\label{eq45a}
& &\epsilon \sum_\mu \int_0^{\infty}\frac{1}{T_l} 
\int_{-s}^{\st} 
G(\vartheta(t_{j,\mu}+\tau)) \\
\nonumber
& &\sum_k  \int_0^\infty \frac{1}{T_l} 
\int_{-s_1}^{\st_1} (F(\vartheta(t_{k,\lambda} + \tau_1))\\
\nonumber
& & + G(\vartheta(t_{k,\lambda} + \tau_1)) \sum_l w_{il} \sum_\rho \kappa(t_{k,\lambda} - t_{l,\rho} + \tau_1))d\tau_1\ (-\Gamma(s_1))ds_1\\
\nonumber
& &\sum_\nu \kappa(t_{j,\mu}-t_{k,\nu}+\tau)\ d\tau\ (-\Gamma(s))ds .
\end{eqnarray}
Here, the first sum extends over all spike times of neuron $j$ with $t-T_l \le t_{j,\mu} \le t$, whereas the second, inner, sum  counts spikes of neuron $k$ with $t_{j,\mu} +\tau  -T_l \le t_{k,\lambda} \le t_{j,\mu} +\tau $. So, the first $G$ records the internal state of neuron $i$ at the time $t_{j,\mu}+\tau$, and the sum over the spike times of $k$ records the weighted input-output correlation coming from $k$ over the time period starting $T_l$ earlier. Of course, we may assume as an approximation that only the weights of selected synapses change whereas the others remain stationary if we wish to analyze a situation where only those specific synapses receive a new input pattern whereas the weights of the others have settled in a stationary state in response to a stationary input pattern. In such a situation, the sum over $k$ then extends only over those neurons whose synapses undergo changes. In particular, we may concentrate on the effect of the synapse from neuron $j$. -- The role of the $\kappa$ terms has already been discussed above.\\

It remains to treat the indirect effect coming from the variation of
$\vartheta$ in response to the variation of $w$. $\vartheta$ satisfies
a differential equation of the form
\bel{eq46}
\dot \vartheta = \Phi(\vartheta) + \Psi(\vartheta) w \kappa
\ee
(thus, comparing this with \rf{eq31}, we have $\dot u = F(\vartheta) +
G(\vartheta) w
\kappa =
\frac{du}{d\vartheta} \dot\vartheta =
\frac{du}{d\vartheta}(\Phi(\vartheta) + \Psi(\vartheta) w
\kappa)$). Differentiating \rf{eq46} with respect to $w$, the
variation $\omega:= \frac{d\vartheta}{dw}$ satisfies
\bel{eq47}
\dot \omega= \Phi^\prime(\vartheta)\omega  +
\Psi^\prime(\vartheta)\omega w\kappa   + \Psi(\vartheta)\kappa.
\ee
Since we wish to understand the effect of learning during some time
interval $[t-h,t]$, we assume that $\omega(t-h)=0$. We then obtain (for $\tau_0 \ge t-h$)
\ba
\label{eq47}
\omega(\tau_0)= & &e^{\int_{t-h}^{\tau_0} (\Phi^\prime(\vartheta(\tau)) +
\Psi^\prime(\vartheta(\tau))w(\tau) \kappa(\tau))d\tau}\\
\nonumber
& &\int_{t-h}^{\tau_0}
\Psi(\vartheta(\tau))\kappa(\tau)e^{-\int_{t-h}^\tau
  (\Phi^\prime(\vartheta(\sigma)) + 
\Psi^\prime(\vartheta(\sigma))w(\sigma)
\kappa(\sigma))d\sigma}d\tau.
\ea
Thus from \rf{eq42b}, the indirect effect is 
\bel{eq48}
(F^\prime(\vartheta)
 + G^\prime(\vartheta)
w\kappa)\frac{d\vartheta}{dw} \dot w,
\ee
with $\frac{d\vartheta}{dw}=\omega$ from
\rf{eq47}. The more precise formula is, after letting h tend to 0,

\begin{eqnarray}
\label{eq49}
& &\epsilon \sum_\mu \int_0^{\infty}\frac{1}{T_l} 
\int_{-s}^{\st}  (F^\prime(\vartheta(t_{j,\mu}+\tau))\\
\nonumber
& &+G^\prime(\vartheta(t_{j,\mu}+\tau)) \sum_m w_{im} \sum_\rho \kappa(t_{j,\mu}+\tau  -t_{m,\rho}))\\
& & \omega(t_{j,\mu}+\tau) \sum_k \dot w_{ik}(t_{j,\mu}+\tau) \ d\tau \ (-\Gamma(s))ds
\nonumber
\end{eqnarray}
with $\omega$ from \rf{eq47} and $\dot w_{ik}$ from \rf{eq32}. 
Assuming $\Psi(\vartheta)$ to be positive, $\frac{d\vartheta}{dw}=\omega$ can be discussed in the same
manner as the direct effect, but the factor in front of it causes a
somewhat different effect. Namely, while we may assume that
$G(\vartheta)$ is positive when $u$ is rising, and negative when it is
falling, the derivative $G^\prime(\vartheta)$ will then naturally be
positive only during the initial phase of the rise of $u$, but then
turn negative to bring $G(\vartheta)$ back to 0 at the firing
value. Thus, here we see a positive effect in response to a weight
change only during that initial phase, but a negative one already
before the neuron fires. The reason for this is of course simply that the strengthening of the synapse will bring the neuron closer to the firing threshold in response to the corresponding input, and thereby push it into a state where it is less receptive to input, or, more precisely, where the input has a comparatively smaller effect. It will depend on the neuron model employed how close to the firing point that effect sets in. Likewise, $F(\vartheta)$ may be negative over part or even all of the range of interest.\\
In any case, we can then insert \rf{eq32} to evaluate the changes in
the input-output correlations explicitly, although the complete
formula will get somewhat complicated. Therefore, we have discussed
here the qualitative effects instead.

\section{Discussion}
We have introduced the learning rule
\bel{eq61}
\dot w_{ij}= \epsilon
\int_{-\infty}^\infty \frac{1}{T_l} \int_{t-T_l}^t \sum_\mu
\delta(\tau+s-t_{j,\mu}) u(\tau) d\tau\ \Gamma(s)ds\\
\ee
where the $t_{j,\mu}$ are the arrival times of spikes from the
presynaptic neuron $j$ and the function $u(t)$ describes the state of
the postsynaptic neuron $i$. $\Gamma(s)$ is the learning window,
positive for negative, negative for positive values of $s$ of
sufficiently small absolute value, and satisfying the symmetry
assumption
\bel{eq62}
\int_{-\infty}^\infty \Gamma(s)ds=0.
\ee
We employ that state function $u$ instead of $i$s spike train because in
general many presynaptic neurons contribute towards a spike of $i$ and
because an analytical treatment needs to incorporate the underlying
rule how the input and the internal dynamics of a neuron generate its
spike pattern whereas we can treat presynaptic spikes as external
events.\\
We could then study for general classes of neuron models how the
relations between incoming spikes and the internal state of $i$
determined the synaptic changes through our learning rule \rf{eq61}.\\
Moreover, we were able to derive an analytical expression for the
converse effect, namely the changes in the input-output relationship
caused by those synaptic weight changes. For that purpose, we needed a
stationarity assumption that without learning, that relation would
have been static. This seems the only reasonable way to isolate the
effects of synaptic learning.\\
As in most theoretical studies on learning through synaptic weight
changes in neural networks, we have exclusively treated excitatory
synapses although inhibiting synapses could also be handled within our framework.\\\\

{\bf Acknowledgement:} I am grateful to the referee for his very
detailed and insightful report that led to substantial improvements of
this paper.


\begin{thebibliography}{100}
\bi{GK} W.Gerstner, W.Kistler, Spiking neuron models, Cambridge
Univ.Press, 2002


\bi{KGvHW}  R.Kempter, W.Gerstner, J.L.van Hemmen, H.Wagner,
Extracting oscillations: Neuronal coincidence detection with noisy
periodic spike input, Neural Comput.10, 1987-2017, 1998

\bi{KGvH1} R.Kempter, W.Gerstner, J.L.van Hemmen, Hebbian learning and
spiking neurons, Phys.Rev.E 59, 4498-4514, 1999

\bi{KGvH2} R.Kempter, W.Gerstner, J.L.van Hemmen, Intrinsic
stabilization of output rates by spike-based learning, 2000

\bi{LKvH} C.Leibold, R.Kempter, J.L.van Hemmen, How spiking neurons
give rise to a temporal-feature map: From synaptic plasticity to
axonal selection, Phys.Rev.E 65, 2002

\bi{vH} J.L.van Hemmen, Theory of synaptic plasticity, in: F.Moss,
S.Gielen (eds.), Handbook of biological physics (Vol.4),
neuro-informatics, neural modelling, Elsevier, 2001, pp.771-823

\bi{XS} X.H.Xie, H.S.Seung, Spike-based learning rules and
stabilization of persistent neural activity, Advances in Neural
Information Processing Systems 12, 199-205, 2000

\bi{MLFS} H.Markram, J.L\"ubke, M.Frotscher, B.Sakmann, Regulation of
synaptic efficacy by coincidence of synaptic APs and EPSPs, Science
275, 213-215, 1997

\bi{Zh} L.Zhang et al, A critical window for cooperation and competition among developing retinotectal synapses, Nature 395, 37-44, 1998

\bi{BP} G.-Q.Bi, M.-M.Poo, Synaptic modification by correlated
activity: Hebb's postulate revisited, Annu.Rev.Neurosci.24, 139-166, 2001
 
\bi{BP2} G.-Q.Bi, M.-M.Poo, Distributed synaptic modification in
neural networks induced by patterned stimulation, Nature 401, 792-796,
1999

\bi{Deb} D.Debanne, B.Gahwiler, S.Thompson, Long-term synaptic plasticity between pairs of individual CA3 pyramidal cells in rat hippocampal slice cultures, J.Physiol. 507, 237-247, 1998

\bi{Feld} D.Feldman, Timing-based LTP and LTD at vertical inputs to layer II/III pyramidal cells in rat barrel cortex, Neuron 27, 45-56, 2000

\bi{FrDan} R.Froemke, Y.Dan, Spike-timing dependent synaptic modification induced by natural spike trains, Nature 416, 433-438, 2002

\bi{BP3} G.-Q.Bi, M.-M.Poo, Activity-induced synaptic modifications in hippocampal culture: dependence on spike timing, synaptic strength, and cell type, J.Neurosc.18 , 10464-10472, 1998

\bi{SMA} S.Song, K.Miller, L.Abbott, Competitive Hebbian learning
through spike-timing dependent synaptic plasticity, Nature
Neuroscience 3, 919-926, 2000

\bi{AN} L.Abbott, S.Nelson, Synaptic plasticity: taming the beast,
Nature Neuroscience Suppl.3, 1178-1183, 2000

\bi{Sjo} P.Sjostrom, G.Turrigiano, S.Nelson, Rate, timing, and
cooperativity jointly determine cortical synaptic plasticity, Neuron
32, 1149-1164, 2001 

\bi{Nis} M.Nishiyama et al., Calcium stores regulate the polarity and input specificity of synaptic modification, Nature 408, 584-588, 2000  

\bi{KD} E.Klann, T.Dever, Biochemical mechanisms for translational regulation in synaptic plasticity, Nature Reviews Neuroscience 5, 931-942, 2004

\bibitem{Er2}   B. Ermentrout, 
              Type I membranes, phase resetting curves, 
              and synchronization, 
              Neural Comp. 8, 979- 1001, 1996


\bibitem{GuEr}  B. Gutkin, B. Ermentrout, 
              Dynamics of membrane excitability determine  
              interspike interval veriability: A link between   
              spike generation mechanisms and cortical spike brain  
              statistics, 
              Neural Comp. 10, 1047- 1065, 1998





\end{thebibliography}
\end{document}